# Distributed Opportunistic Channel Access in Wireless Relay Networks

Zhou Zhang and Hai Jiang

University of Alberta, Canada

Email: zzhang6@ualberta.ca, hai.jiang@ece.ualberta.ca

**Abstract**

In this paper, the problem of distributed opportunistic channel access in wireless relaying is investigated. A relay network with multiple source-destination pairs and multiple relays is considered. All the source nodes contend through a random access procedure. A winner source node may give up its transmission opportunity if its link quality is poor. In this research, we apply the optimal stopping theory to analyze when a winner node should give up its transmission opportunity. By assuming the winner node has information of channel gains of links from itself to relays and from relays to its destination, the existence and uniqueness of an optimal stopping rule are rigorously proved. It is also found that the optimal stopping rule is a pure-threshold strategy. The case when the winner node does not have information of channel gains of links from relays to its destination is also studied. Two stopping problems exist, one in the main layer (for channel access of source nodes), and the other in the sub-layer (for channel access of relay nodes). An intuitive stopping rule, where the sub-layer and the main layer maximize their throughput respectively, is shown to be a semi-pure-threshold strategy. The intuitive stopping rule turns out to be non-optimal. An optimal stopping rule is then derived theoretically. Our research reveals that multi-user (including multi-source and multi-relay) diversity and time diversity can be fully utilized in a relay network by our proposed strategies.

*Keywords* – Relay, opportunistic channel access, optimal stopping.

## I. INTRODUCTION

In a wireless network, the wireless medium is shared by all users. The medium access control (MAC) layer is to coordinate the channel access of the users in a orderly and efficient manner. However, since the link between a transceiver pair usually experiences fading and/or shadowing, it is now well recognized that the MAC layer protocol should be jointly designed with the physical layer. This has led to a cross-layer design concept, namely, channel-aware scheduling or opportunistic channel access. The basic idea is to let the MAC layer aware of the physical layer states. If a transmitter finds that its channel quality is poor, it may give up or be asked to give up (by a central controller, such as the base station in a cellular network) its channel access opportunity, with an expectation that there are other users with good channel quality[1] and those users can utilize its channel access opportunity and gain more. Although the user may sacrifice in a short term, it will get more in the long term, because later the user, when with good channel condition, may utilize the channel access opportunity of other users that have poor channel quality.

---

[1]This expectation is reasonable since the users' channels are independent.

Opportunistic channel access has received much attention in the literature, particularly in centralized networks [1], [2]. A central controller can collect the channel quality information of the users, and schedule only those users with the best channel conditions. On the other hand, the research on distributed opportunistic channel access is still in its infancy. Without a central controller, it is hard for a user to decide when to give up its transmission opportunities. An intuitive way is to categorize the channel of a user into two states: good state when the channel gain is above a threshold; and bad state otherwise. Then a user gives up its channel access opportunity when its channel is bad. Apparently the multi-user diversity (i.e., different users experience different channel gains) and time diversity (i.e., a user experiences different channel gain when time varies) are not fully utilized by the intuitive method. This problem was address recently in [3], by means of optimal stopping. The major idea is to let all the users contend for channel access. It is found that, 1) if the winner in a contention has an achievable (transmission) rate smaller than a threshold (which can be obtained by solving an equation), it is optimal for the winner to give up its transmission opportunity and all users continue to contend again; and 2) if the winner in a contention has an achievable rate larger than the threshold, it is optimal for the winner to *stop* here, i.e., to utilize the transmission opportunity and transmit its data. The beautiful part of the work is in the *pure-threshold strategy*, which is easy to implement. As extensions to the work in [3], interference channel which can tolerate multiple users transmitting is considered in [4] where more than one nodes can share the channel simultaneously, and delay constraints are considered for real-time service in [5]. Pure-threshold strategies are also derived in [4], [5].

In this paper, we investigate opportunistic channel access in a relay network, since wireless relaying has recently attracted a lot of research interests [6]–[16]. We focus on distributed opportunistic channel access in relay networks. We consider the scenario of multiple source-destination pairs aided by multiple relays. Since each source-destination pair involves two hops: from source to relays and from relays to the destination, the problem of opportunistic channel access in relay networks is quite different from those in a single-hop network (e.g., in reference [3]–[5]), and is challenging as multi-source diversity, multi-relay diversity, and time diversity should be all exploited. Two scenarios are considered: when the a source node has information of channel gains of the links from itself to relays and from relays to its destination, and when a source node only has information of channel gains of the links from itself to relays. In the first scenario, it is found that a pure-threshold strategy uniquely exists to optimize the average system throughput. There are two stopping problems in the second scenario, one in the main layer (for channel access of source nodes) and the other in the sub-layer (for channel access of relay nodes). An intuitive strategy is proposed, which is shown to be a semi-pure-threshold strategy. However, the strategy is not optimal. We also theoretically derive an optimal strategy for the second scenario. The rest of this paper is organized as follows. System model is introduced in Section II. The system throughput maximization problem in the two scenarios are theoretically solved in Section III and IV, respectively, followed by concluding remarks in Section V.

## II. SYSTEM MODEL

Consider $K$ source-destination pairs aided by $L$ relays. Amplify-and-forward (AF) mode [7], [11], [17], [18] is adopted when a relay is helping a source-destination pair. The transmission power of a source node and a relay node is $P_s$ and $P_r$, respectively.

We use a similar random access procedure to that in [3]. In a slot with duration $\tau$, assume each source contends for channel access with probability $p_0$. Then in a slot, the contention is successful if there is only one contender, with probability $p_s = Kp_0(1-p_0)^{K-1}$. Therefore, the number of slots, $K_s$, until a successful contention follows a geometric distribution with $\text{Prob}(K_s = n) = p_s(1-p_s)^{n-1}$. Then the duration of total contentions until a successful contention is $\tau K_s$ with its expectation as $\frac{\tau}{p_s}$.

If the $i$th source node wins a channel contention and transmits, and relay $j$ helps forward its traffic to the $i$th destination, the maximal rate that can be achieved in AF mode is

$$\log_2(1 + \frac{P_s P_r |f_{ij}|^2 |g_{ji}|^2}{1 + P_s |f_{ij}|^2 + P_r |g_{ji}|^2}) \tag{1}$$

where $f_{ij}$ and $g_{ji}$ are channel gains from the $i$th source to the $j$th relay and from the $j$th relay to the $i$th destination. Assume the channel gains from a source to a relay and from a relay to a destination follow a complex Gaussian distribution with mean being zero and variance being $\sigma_f^2$ and $\sigma_g^2$, respectively.

In this research we aim at distributed opportunistic channel access with the help of optimal stopping theory. Some notations are defined as follows. After the previous data transmission, an *observation* is the process of channel contention until a successful contention (i.e., a winner appears). After each observation, the winner decides whether to continue a new observation (i.e., a new contention round is started) or to stop observation (i.e., the user transmits its data). In the $n$th observation, let $s(n)$ denote the contention winner, $k(n)$ denote the time spent in the observation (equal to $\tau$ times the number of slots used in the contention), $f_{s(n)i}$ ($i = 1, ..., L$) denote the channel gain between source node $s(n)$ and the $i$th relay, and $g_{is(n)}$ denote the channel gain between the $i$th relay and destination node $s(n)$. In other words, the observed information in the $n$th observation is: $X(n) := \{s(n), k(n), f_{s(n)1}(n), ..., f_{s(n)L}(n), g_{1s(n)}(n), ..., g_{Ls(n)}(n)\}$. For the $n$th observation, the reward $y_n$ is the total traffic volume that can be sent if the winner transmits its data traffic, which is a function of $X(n)$, and the cost $T_n$ is the total waiting time from the first observation until the $n$th observation plus the data transmission time. If the procedure decides to stop at the $N$th observation, then the average system throughput is $\frac{Y_N}{T_N}$. In the sequel, capital $N$ is called the stopping time. And our objective is to find the optimal stopping time (also called optimal stopping rule), $N^*$, which attains average system throughput $\sup_{N \geq 0} \frac{E(Y_N)}{E(T_N)}$. Here $E(\cdot)$ means expectation. According to [19, Chapter 6], this maximal-expected-return problem can be equivalently transformed into a standard form with its reward changed to be $Y_N - \lambda^* T_N$. In particular, to get $N^*$, we need to find an optimal rule to reach maximal expected reward

$$V^*(\lambda^*) = E(Y_N) - \lambda^* E(T_N) \tag{2}$$

where $\lambda^*$ satisfies $\sup_{N\geq 0}\{E(Y_N) - \lambda^* E(T_N)\} = 0$. Here $\lambda^*$ is actually the maximal system throughput in our problem. This transformation method will be used when we solve the optimal stopping problems in our research, as shown in the subsequent sections.

Assume that the winner in the $n$th observation (i.e., source node $s(n)$) has information of channel gains $\{f_{s(n)1}(n), ..., f_{s(n)L}(n), g_{1s(n)}(n), ..., g_{Ls(n)}(n)\}$.[2] If the winner $s(n)$ decides to stop, it selects the relay that renders the maximal source-to-destination rate, given as

$$R_n = \max_{j\in\{1,...,L\}} \left\{\log_2(1 + \frac{P_s P_r |f_{s(n)j}(n)|^2 |g_{js(n)}(n)|^2}{1 + P_s|f_{s(n)j}(n)|^2 + P_r|g_{js(n)}(n)|^2})\right\}. \quad (3)$$

The total transmission time is $T$, where the transmission time from source $s(n)$ to the selected relay and from the selected relay to destination $s(n)$ are both $T/2$.

## III. AVERAGE THROUGHPUT MAXIMIZATION BY OPTIMAL STOPPING RULE

To formulate our research problem as an optimal stopping problem, in the $n$th observation, the reward is $Y_n = \frac{T}{2}R_n$ with the spent time denoted as $T_n = \tau \sum_{i=1}^{n} K_i + T$. For finding a strategy $N^*$ to control source nodes' random access and finally achieve maximal average system throughput $\frac{E(Y_N)}{E(T_N)}$, it is equivalent [19] to design a rule which attains $\sup_{N\geq 0}\left\{\frac{T}{2}E(R_N) - \lambda^* E\left(T + \tau \sum_{i=1}^{N} K_i\right)\right\}$ where $\lambda^*$ satisfies

$$\sup_{N\geq 0}\left\{\frac{T}{2}E(R_N) - \lambda^* E\left(T + \tau \sum_{i=1}^{N} K_i\right)\right\} = 0.$$

Before giving the optimal stopping rule $N^*$, two conditions $E\left\{\sup_n\left(\frac{T}{2}R_n - \lambda T + \tau \sum_{i=1}^{n} K_i\right)\right\} < \infty$ and $\limsup_{n\to\infty}\left\{\frac{T}{2}R_n - \lambda\left(T + \tau \sum_{i=1}^{n} K_i\right)\right\} = -\infty$ a.s. should be checked which guarantee the existence of an optimal stopping rule. Here $\lambda$ can be viewed as the system throughput, while $\lambda^*$ has the physical meaning of maximal system throughput.

*Lemma 1:* let $c > 0$, we have $E(\frac{T}{2}R_n) < \infty$ and $E\left[\sup_n(\frac{T}{2}R_n - nc)\right] < \infty$.

Proof: See Appendix I.

By decomposition similar to (43) in [3] and using Lemma 1, the first condition for existence of an optimal stopping rule can be proved.

---

[2]The case when source node $s(n)$ has only information $\{f_{s(n)1}(n), ..., f_{s(n)L}(n)\}$ will be discussed in Section IV.

*Lemma 2:* The second condition is also satisfied, namely $\limsup_{n\to\infty} \left[\frac{T}{2}R_n - \lambda\left(T + \tau\sum_{i=1}^{n}K_i\right)\right] = -\infty$ a.s.

Proof: See Appendix II.

Based on Lemma 1 and Lemma 2, the existence of an optimal stopping rule is guaranteed.

*Theorem 1:* The optimal stopping rule which achieves maximal system throughput $\sup_{N\geq 0}\frac{E(Y_N)}{E(T_N)}$ is given as follows: $N^* = \min\{n \geq 1 : R_n \geq 2\lambda^*\}$ where $\lambda^*$ is the solution of the equation $E\left\{\max\left\{\frac{T}{2}R_n - \lambda T, 0\right\}\right\} = \frac{\lambda\tau}{p_s}$.

Proof: See Appendix III.

With threshold $2\lambda^*$ as a constant, our derived rule $N^*$ has a pure-threshold structure and achieves the maximal system throughput $\lambda^* = \frac{E(Y_{N^*})}{E(T_{N^*})}$.

*Theorem 2:* As the solution of the equation $E\left\{\max\left\{\frac{T}{2}R_n - \lambda T, 0\right\}\right\} = \frac{\lambda\tau}{p_s}$, the maximal system throughput $\lambda^*$ always uniquely exists.

Proof: See appendix IV.

The uniqueness of $\lambda^*$ is consistent with its physical meaning as the optimal system throughput.

For the pure-threshold rule $N^* = \min\{n \geq 1 : R_n \geq 2\lambda^*\}$, we can acquire following conclusion.

*Corollary 1:* With $\{R_n\}_{n=1,\ldots,\infty}$ i.i.d. and pure-threshold structure of $N^*$, the stopping time denoted $N$ determined by the optimal stopping rule $N^*$ follows a geometric distribution with $\text{Prob}(N = n) = F_R(2\lambda^*)^{n-1}(1 - F_R(2\lambda^*))$ where $F_R$ means cumulative distribution function (CDF) of $R_n$. Let $R_{N^*}$ denote the achievable rate at the stopped time. It has the CDF as $F_{R_{N^*}}(x) = I(x \geq 2\lambda^*)\frac{F_R(x) - F_R(2\lambda^*)}{1 - F_R(2\lambda^*)}$ where $I(\cdot)$ means an indicator function.

*Corollary 2:* With the stopping time $N$ determined by the rule $N^*$ geometrically distributed, the expectation of the stopping time $E(N) = \frac{1}{1 - F_R(2\lambda^*)}$ is finite. With $E(\tau K_n) = \frac{\tau}{p_s}$, according to wald theorem [19] we have $E(T_N) = E(\tau K_n)E(N) < \infty$.

In fact, these results arise from the pure-threshold structure of the optimal stopping rule $N^*$. In addition, the pure-threshold structure largely simplifies our strategy realization. In details, after the $n$th successful channel contention, source node $s(n)$ wins the channel and calculates its achievable transmission rate $R_n$ (which is via the best relay). If $R_n \geq 2\lambda^*$, source node $s(n)$ transmits to the best relay node and the best relay node helps forward to destination node $s(n)$; otherwise, source node $s(n)$ gives up the transmission opportunity and re-contends the channel with the other $K-1$ source nodes again. In this way, the maximal average system throughput $\lambda^*$ can be achieved.

## IV. CASE WHEN A SOURCE NODE DOES NOT HAVE INFORMATION OF CHANNEL GAINS IN THE SECOND HOP

In previous sections, the winner source node in each observation has the information of channel gains of the links from itself to relays and from relays to its destination. Next we consider a more practical case that the winner source node in each observation only has channel gain information of links from itself to relays. The channel access procedure includes two parts: from sources to relays, and from relays to destinations. In the first part of the channel access, in a slot (with duration $\tau/2$), each source node contends for channel access with probability $p_0$. Then in a slot, the contention is successful if there is only one contender, with probability $p_s = Kp_0(1-p_0)^{K-1}$. Upon a successful contention at the $n$th observation, the winner (a source node), denoted $s(n)$, decides whether to stop and transmit, or to give up its transmission opportunity and start a new contention with other source nodes. If the winner decides to stop, it broadcasts its data to all relays. In the broadcast, the channel gain information of its links to all relays is also included. Then the second part of channel access procedure starts. In a slot (with duration $\tau/2$), each relay node contends for channel access with probability $p_1$. Then in a slot, the contention is successful if there is only one contender, with probability $p_r = Lp_1(1-p_1)^{L-1}$. Upon a successful contention at the $m$th observation, the winner (a relay node), denoted $s(m)$, decides whether to stop and transmit, or to give up its transmission opportunity and start a new contention with other relay nodes. If the winner decides to stop, it forwards its data (received from the source node $s(n)$) to the corresponding destination.

In the first or second part of the channel access procedure, the number of slots spent for a successful source or relay contention is geometrically distributed, with expectation $\frac{\tau}{2p_s}$ or $\frac{\tau}{2p_r}$, respectively.

It can be seen that the channel access is actually a bi-level stopping problem: the main layer for channel access of source nodes, and the sub-layer for channel access of relay nodes. In the main layer, the source nodes do not have channel gain information of links in the second hop (from relays to destinations). However, it is assumed the source nodes have statistical information (e.g., channel gain distribution) of channel gains in the second hop. Therefore, in the main layer, the reward (which is the source-to-destination data volume) in the $n$th observation is the expected reward in the sub-layer. On the other hand, in the sub-layer, the channel gain information of links from the winner to relays is already known. In other words, the sub-layer stopping problem should be based on channel gain information of the links in the first hop. It is also assumed that, in the sub-layer, a winner in a successful contention can have its channel gain information to its destination.

In the main layer, let $n$ and $N$ denote the observation index and stopping time, respectively. And in the sub-layer, let $m$ and $M$ denote the observation index and stopping time, respectively.

### A. Intuitive Stopping Rule

An intuitive method to solve the bi-level stopping problem is to let the sub-layer and main layer apply optimal stopping theory to maximize sub-layer and main-layer throughput, respectively.

In the sub-layer, the relays already know the information of $\{f_{s(n)1}(n), ..., f_{s(n)L}(n)\}$. Then in the $m$th observation, the achievable rate of the winner, $s(m)$, is

$$R_m = \left\{\log_2(1 + \frac{P_s P_r |f_{s(n)s(m)}(n)|^2 |g_{s(m)s(n)}(m)|^2}{1 + P_s |f_{s(n)s(m)}(n)|^2 + P_r |g_{s(m)s(n)}(m)|^2})\right\}. \quad (4)$$

And the reward in the $m$th observation is $Y_m = \frac{T}{2} R_m$. Then we need to find the the optimal stopping rule $M^*$ in the sub-layer to attain the maximal $\lambda^* = \sup_{M \geq 0} \frac{E(Y_M)}{E(T_M)}$.

In the main layer, if the stopping time is $N$, then the reward is $E(Y_{M^*})$, and the waiting time is $E(T_{M^*}) + T_N$. Then we need to find the optimal stopping rule $N^*$ to attain the maximal $\sup_{N \geq 0} \frac{E(E(Y_{M^*}))}{E(E(T_{M^*}) + T_N)}$. Note that in the numerator or denominator in $\frac{E(E(Y_{M^*}))}{E(E(T_{M^*}) + T_N)}$, the inner expectation is for the second hop, while the outer expectation is for the first hop.

For the sub-layer optimal stopping problem, we have the following theorem.

*Theorem 3:* Conditioned on a main-layer channel gain realization $\{f_{s(n)1}(n), ..., f_{s(n)L}(n)\}$, a sub-layer optimal stopping rule achieving the maximal sub-layer throughput $\lambda^* = \sup_{M \geq 0} \frac{E(Y_M)}{E(T_M)}$ is given as this form: $M^* = \min\{m \geq 1 : R_m \geq \lambda^*\}$ where $\lambda^*$ is the unique solution of the equation $E\{\max\{R_m - \lambda, 0\}\} = \frac{\lambda\tau}{Tp_r}$ and always exists.

Proof: See appendix V.

The sub-layer optimal stopping rule has the following property.

*Corollary 3:* With channel gains $\{f_{s(n)1}(n), ..., f_{s(n)L}(n)\}$ having finite values, $0 \leq \lambda^* < \infty$, $E(T_{M^*}) < \infty$ and $E(Y_{M^*}) < \infty$.

Proof: See Appendix VI.

Based on the acquired strategy $M^*$ for the sub-layer stopping problem, a main-layer optimal stopping rule which achieves maximal system throughput is given in the following theorem.

*Theorem 4:* An optimal stopping rule for the main-layer problem is of the form $N^* = \min\{n \geq 1 : R_n^1 - \gamma^* R_n^2 \geq \gamma^* \frac{T}{2}\}$ where $\gamma^*$ satisfies the equation $E\{\max\{R_n^1 - \gamma R_n^2 - \gamma \frac{T}{2}, 0\}\} = \frac{\gamma\tau}{2p_s}$, $R_n^1 = \lambda^* E(T_{M^*})$ and $R_n^2 = E(T_{M^*})$.[3]

Proof: See appendix VII. Note that here $\gamma^*$ is actually the maximal main-layer system throughput.

From Theorem 3 and 4 it can be seen that, the intuitive optimal stopping rule $\{N^*, M^*\}$ with $M^* = \min\{m \geq 1 : R_m \geq \lambda^*\}$ and $N^* = \min\{n \geq 1 : R_n^1 - \gamma^* R_n^2 \geq \gamma^* \frac{T}{2}\}$ has semi-pure-threshold structure. In details, with sub-layer stopping rule $M^*$, its threshold is not a constant but a function of channel gains in the first hop. Different from $M^*$, the main-layer stopping rule $N^*$ has a constant threshold. The intuitive stopping rule can be implemented as follows.

For source node channel access, upon a successful contention, the winner $s(n)$ has the information of its channel gains $\{f_{s(n)1}(n), ..., f_{s(n)L}(n)\}$. The winner can calculate $R_n^1$ and $R_n^2$ by solving the sub-layer optimal stopping problem conditioned on channel gains in the first hop as $\{f_{s(n)1}(n), ..., f_{s(n)L}(n)\}$. For the source node channel access, $\gamma^*$ is a constant satisfying $E\{\max\{R_n^1 - \gamma R_n^2 - \gamma \frac{T}{2}, 0\}\} = \frac{\gamma\tau}{2p_s}$.

- If $R_n^1 - \gamma^* R_n^2 < \gamma^* \frac{T}{2}$, the source node gives up its transmission opportunity and re-contend with other source nodes.

---
[3]Note that here $M^*$ is the optimal stopping rule of the sub-layer conditioned on channel gains in the first hop as $\{f_{s(n)1}(n), ..., f_{s(n)L}(n)\}$, and $\lambda^*$ is the corresponding maximal throughput in the sub-layer stopping problem.

- If $R_n^1 - \gamma^* R_n^2 \geq \gamma^* \frac{T}{2}$, the source node broadcasts its data to all relays, and the relay node channel access starts. Upon a successful contention of a relay node $s(m)$ in the $m$th observation, for the particular channel gains in the first hop as $\{f_{s(n)1}(n), ..., f_{s(n)L}(n)\}$, relay node $s(m)$ calculates $\lambda^*$ by solving $E\{\max\{R_m - \lambda, 0\}\} = \frac{\lambda\tau}{T_{pr}}$. Then if $R_m < \lambda^*$, the relay node $s(m)$ gives up its transmission opportunity, and re-contend with other relay nodes; otherwise, the relay node $s(m)$ forwards its received data (from source node $s(n)$) to destination node $s(n)$, and the transmission process for the packet from $s(n)$ is complete, and all source nodes start a new round of contention again.

## B. Non-optimality of Intuitive Stopping Rule

The intuitive stopping rules $\{N^*, M^*\}$ first maximizes sub-layer system throughput and then maximizes that of main-layer system. It is interesting to notice that the intuitive stopping rule is not optimal, as follows.

The expected system throughput can be expressed as $\frac{E\{[\lambda^* E(T_{M^*})]_{N^*}\}}{E\{[E(T_{M^*})]_{N^*} + T_{N^*}\}}$ in the intuitive stopping rule, where the subscript $N^*$ in $[\lambda^* E(T_{M^*})]_{N^*}$ and $[E(T_{M^*})]_{N^*}$ means the optimal stopping rule in the sub-layer is conditioned on the channel gains in the first hop when the main layer follows the stopping rule $N^*$. The sub-layer stopping rule $M^*$ maximizes $\frac{\lambda^* E(T_{M^*})}{E(T_{M^*})}$. Considering the term $T_{N^*}$ in the expression of the expected system throughput, the sub-layer stopping rule $M^*$, which maximizes $\frac{\lambda^* E(T_{M^*})}{E(T_{M^*})}$, may not maximize $\frac{E\{[\lambda^* E(T_{M^*})]_{N^*}\}}{E\{[E(T_{M^*})]_{N^*} + T_{N^*}\}}$. For example, for a sub-layer non-optimal stopping rule $M^\dagger$, we have $\lambda^\dagger < \lambda^*$. However, if $[E(T_{M^\dagger})]_{N^*} < [E(T_{M^*})]_{N^*}$, it is possible to have $\frac{[\lambda^* E(T_{M^*})]_{N^*}}{[E(T_{M^*})]_{N^*} + T_{N^*}} < \frac{[\lambda^\dagger E(T_{M^\dagger})]_{N^*}}{[E(T_{M^\dagger})]_{N^*} + T_{N^*}}$.

## C. Proposed Optimal Stopping Rule

From the previous discussion, $M^*$ in the intuitive stopping rule maximizes the sub-layer system throughput, not the main layer system throughput. Therefore, in the proposed optimal stopping rule, we do not let the sub-layer maximize the sub-layer system throughput. Rather, we let the sub-layer achieve maximal average award $\sup_{M \geq 0} E(Y_M - \gamma T_M)$, where $\gamma > 0$ represents the throughput in the main layer.

Since we have two stopping problems, we use $V^*$ and $W^*$ to denote the maximal expected reward after the problem transformation (similar to (2)) in the main layer and sub-layer, respectively.

For the sub-layer stopping problem, we have the following theorem.

*Theorem 5:* For fixed $\gamma \geq 0$, the optimal stopping rule $M^*(\gamma)$ for maximizing $E(Y_M - \gamma T_M)$ is of the form: $M(\gamma) = \min\left\{m \geq 1 : \frac{T}{2}R_m \geq W^*(\gamma) + \frac{T}{2}\gamma\right\}$ where $W^*(\gamma)$ satisfies the equations:

$$E\left[\max\left(\frac{T}{2}R_m - \frac{T}{2}\gamma, W^*(\gamma)\right)\right] = W^*(\gamma) + \frac{\gamma\tau}{2p_r} \tag{5}$$

Proof: See Appendix VIII.

Although Theorem 5 is for any particular value of $\gamma$, it is desired the sub-layer stopping rule is corresponding to the maximal system throughput $\gamma^*$. How to obtain the value of $\lambda^*$ will be discussed shortly. Therefore, for the main-layer stopping problem, it is assumed that the sub-layer stopping problem follows the rule $M^*(\gamma^*)$, and we have the following theorem for the main-layer stopping problem, the objective of which is to achieve the maximal system throughput.

*Theorem 6:* With the sub-layer system following the strategy $M^*(\gamma^*)$, an optimal strategy to maximize the average system throughput is given as $N^* = \min\left\{n \geq 1 : W^*(\gamma^*) \geq \frac{T}{2}\gamma^*\right\}$ where $\gamma^*$ satisfies $E\left[\max(W^*(\gamma) - \frac{T}{2}\gamma, 0)\right] = \frac{\tau\gamma}{2p_s}$.

Proof: See Appendix IX.

Based on Theorem 5 and 6, we can see that our proposed stopping rule $\{N^*, M^*\}$ has the form of $M(\gamma^*) = \min\left\{m \geq 1 : \frac{T}{2}R_m \geq W^*(\gamma^*) + \frac{T}{2}\gamma^*\right\}$ and $N^* = \left\{n \geq 1 : W^*(\gamma^*) \geq \frac{T}{2}\gamma^*\right\}$, which achieves average system throughput maximum $\gamma^*$. Here $\gamma^*$ is a constant satisfying

$$E\left[\max(W^*(\gamma) - \frac{T}{2}\gamma, 0)\right] = \frac{\tau\gamma}{2p_s}$$

where $W^*(\gamma)$ is unique root of $E\left[\max\left(\frac{T}{2}R_m - \frac{T}{2}\gamma, W^*(\gamma)\right)\right] = W^*(\gamma) + \frac{\gamma\tau}{2p_r}$. Therefore, value of constant $\gamma^*$ can be obtained numerically.

Note that the proposed rule $\{N^*, M^*\}$ has also semi-pure-threshold structure, as in the main layer, the threshold is a constant, while in the sub-layer, the threshold $W^*(\gamma^*)$ is conditioned on the channel gain realization in the first hop.

The proposed stopping rule can be carried out as follows. Here we assume each node knows the value of constant $\gamma^*$.

For source node channel access, upon a successful contention, the winner $s(n)$ has the information of its channel gains $\{f_{s(n)1}(n), ..., f_{s(n)L}(n)\}$. The winner can calculate $W^*(\gamma^*)$ by solving the sub-layer optimal stopping problem conditioned on channel gains in the first hop as $\{f_{s(n)1}(n), ..., f_{s(n)L}(n)\}$.

- If $W^*(\gamma^*) < \frac{T}{2}\gamma^*$, the source node gives up its transmission opportunity and re-contend with other source nodes.

- If $W^*(\gamma^*) \geq \frac{T}{2}\gamma^*$, the source node broadcasts its data to all relays, and the relay node channel access starts. Upon a successful contention of a relay node $s(m)$, for the particular channel gains in the first hop as $\{f_{s(n)1}(n), ..., f_{s(n)L}(n)\}$, relay node $s(m)$ calculates $W^*(\gamma^*)$ by solving $E\left[\max\left(\frac{T}{2}R_m - \frac{T}{2}\gamma^*, W^*(\gamma^*)\right)\right] = W^*(\gamma^*) + \frac{\gamma^*\tau}{2p_r}$. If $\frac{T}{2}R_m < W^*(\gamma^*) + \frac{T}{2}\gamma^*$, the relay node $s(m)$ gives up its transmission opportunity, and re-contend with other relay nodes; otherwise, the relay node $s(m)$ forwards its received data (from source node $s(n)$) to the destination $s(n)$, and the transmission process for the packet from $s(n)$ is complete, and all source nodes start a new round of contention again.

### D. Optimality of Proposed Stopping Rule

According to Lemma 6.1 and Theorem 6.1 in [19], we have the following properties of average reward $V^*(\gamma)$.

*Lemma 3:* For $\gamma \geq 0$, the function $V^*(\gamma) := \sup_{N \geq 0} \{E\{[E(Y_M)]_N - \gamma\{[E(T_M)]_N + T_N\}\}\}$ is decreasing and convex function of $\gamma$.

*Lemma 4:* For some $\gamma$, $V^*(\gamma) := \sup_{N \geq 0} \{E\{[E(Y_M)]_N - \gamma\{[E(T_M)]_N + T_N\}\}\} = 0$ if and only if $\gamma = \sup_{N \geq 0} \left\{\frac{E\{[E(Y_M)]_N\}}{E\{\{[E(T_M)]_N + T_N\}\}}\right\}$.

Remark: By combing these two lemmas, for a fixed sub-layer stopping rule $M$ we can find an optimal stopping rule $N^\dagger$ which attains $\sup_{N \geq 0} \left\{\frac{E\{[E(Y_M)]_N\}}{E\{\{[E(T_M)]_N + T_N\}\}}\right\}$. The rule $N^\dagger$ is obtained by solving an optimal stopping problem with reward defined as $[E(Y_M)]_N - \gamma^\dagger\{[E(T_M)]_N + T_N\}$ where $\gamma^\dagger = \sup_{N \geq 0} \left\{\frac{E\{[E(Y_M)]_N\}}{E\{\{[E(T_M)]_N + T_N\}\}}\right\}$. In other words, to prove the optimality of our proposed stopping rule ($N^*$, $M^*$), we need only to prove that $M^*$ can maximize $V^*(\gamma)$, as shown in the following lemma.

*Lemma 5:* For $\gamma \geq 0$, the function $V^*(\gamma) := \sup_{N \geq 0} \{E\{[E(Y_M)]_N - \gamma\{[E(T_M)]_N + T_N\}\}\}$ achieves its maximum by following the stopping rule $M^*$. The rule $M^*$ is an optimal stopping rule providing maximal expected reward $E(Y_M - \gamma T_M)$ in the sub-layer.

*Proof:* Since
$$V^*(\gamma) : = \sup_{N \geq 0} \{E\{[E(Y_M)]_N - \gamma\{[E(T_M)]_N + T_N\}\}\}$$
$$= \sup_{N \geq 0} \{E\{[E(Y_M) - \gamma E(T_M)]_N - \gamma T_N\}\}$$

for $\gamma \geq 0$, by using the stopping rule $M^*$ to maximize the average reward $E(Y_M - \gamma T_M)$ in the sub-layer, $V^*(\gamma)$ can be maximized. ∎

With sub-layer problem solved by $M^*$, according to our remark of Lemma 3 and Lemma 4, an optimal stopping rule $N^*$ attains maximal average system throughput $\gamma^* = \sup_{N \geq 0} \left\{ \frac{E\{[E(Y_{M^*})]_N\}}{E\{\{[E(T_{M^*})]_N + T_N\}\}} \right\}$. This completes the proof of optimality of our proposed stopping rule $\{N^*, M^*\}$.

## V. CONCLUSION

In a wireless relay network, the source nodes and relays nodes all experience independent fading. It is desired to fully exploit the multi-source diversity, multi-relay diversity, and time diversity. To achieve this, opportunistic channel access is needed, which is investigated in our research in a distributed structure. For the two considered scenarios (with source nodes having or not having channel state information in the second hop), we derived the optimal rules for opportunistic channel access. This research should provide insight to the design of channel-aware MAC protocols in wireless relay network. Further research may include the cases with limited channel state information and with quality-of-service constraints.

## APPENDIX I
## PROOF OF LEMMA 1

Achievable transmission rate at the $n$th observation and its expectation can be re-written as

$$R_n = \sum_{i=1}^{K} \left\{ I([s(n) = i]) \max_{j \in \{1,...,L\}} \left\{ \log_2(1 + \frac{P_s P_r |f_{ij}(n)|^2 |g_{ji}(n)|^2}{1 + P_s |f_{ij}(n)|^2 + P_r |g_{ji}(n)|^2}) \right\} \right\} \quad (6)$$

$$E(R_n) = \sum_{i=1}^{K} \left\{ \frac{1}{K} E \left\{ \max_{j \in \{1,...,L\}} \left\{ \log_2(1 + \frac{P_s P_r |f_{ij}(n)|^2 |g_{ji}(n)|^2}{1 + P_s |f_{ij}(n)|^2 + P_r |g_{ji}(n)|^2}) \right\} \right\} \right\} \quad (7)$$

where $I(\cdot)$ means an indicator function.

Since $f_{ij}$ and $g_{ji}$ follow complex Gaussian distribution with mean being zero and variance being $\sigma_f^2$ and $\sigma_g^2$, respectively, we have $E(|f_{ij}|^2) = \delta_f^2$ and $E(|g_{ji}|^2) = \delta_g^2$. Then we have

$$\begin{aligned}
E(R_n) &= \sum_{i=1}^{K} \left\{ \frac{1}{K} E \left\{ \max_{j \in \{1,\ldots,L\}} \left\{ \log_2(1 + \frac{P_s P_r |f_{ij}(n)|^2 |g_{ji}(n)|^2}{1 + P_s |f_{ij}(n)|^2 + P_r |g_{ji}(n)|^2}) \right\} \right\} \right\} \\
&< \sum_{i=1}^{K} \left\{ \frac{1}{K} E \left\{ \sum_{j=1}^{L} \left\{ \log_2(1 + \frac{P_s P_r |f_{ij}(n)|^2 |g_{ji}(n)|^2}{1 + P_s |f_{ij}(n)|^2 + P_r |g_{ji}(n)|^2}) \right\} \right\} \right\} \\
&\stackrel{(a)}{\leq} \sum_{i=1}^{K} \left\{ \frac{1}{K} \left\{ \sum_{j=1}^{L} \frac{E(P_s |f_{ij}|^2) E(P_r |g_{ji}|^2)}{\ln 2} \right\} \right\} \\
&= \frac{1}{\ln 2} L P_s P_r \sigma_f^2 \sigma_g^2 < \infty
\end{aligned} \qquad (8)$$

where (a) comes from the fact that for $x, y \geq 0$, we have

$$\log_2(1 + \frac{xy}{1+x+y}) \leq \frac{\frac{xy}{1+x+y}}{\ln 2} \leq \frac{xy}{\ln 2}. \qquad (9)$$

Based on [19, Theorem 4.1], from $E(R_n) < \infty$, we have $\sup_n \left(\frac{T}{2} R_n - nc\right) < \infty$ a.s., which leads to $E\left\{\sup_n \left(\frac{T}{2} R_n - nc\right)\right\} < \infty$.

## APPENDIX II
## PROOF OF LEMMA 2

Using a similar method in [3], for $0 < \varepsilon < E(K_i) = \frac{1}{p_s}$, we have the following decomposition:

$$\frac{T}{2} R_n - \lambda(T + \tau \sum_{i=1}^{n} K_i) = \left[\frac{T}{2} R_n - n\lambda\tau\left(\frac{1}{p_s} - \varepsilon\right) - T\lambda\right] + \left[\lambda\tau \sum_{i=1}^{n} \left(\frac{1}{p_s} - \varepsilon - K_i\right)\right]. \qquad (10)$$

From [19, Theorem 4.1], $\frac{1}{p_s} - \varepsilon > 0$, and the result $E(R_n) < \infty$ (as shown in proof of Lemma 1), we obtain that

$$\lim_{n \to \infty} \left[\frac{T}{2} R_n - n\lambda\tau\left(\frac{1}{p_s} - \varepsilon\right)\right] = -\infty \text{ a.s.} \qquad (11)$$

Next we focus on the second component on the right-hand side of (10).

Using [19, Theorem 4.2], when $E\left(\frac{1}{p_s} - \varepsilon - K_i\right) < 0$ holds, $E\left(\sup_{n \geq 0} \sum_{i=1}^{n} \left(\frac{1}{p_s} - \varepsilon - K_i\right)\right) < \infty$ if and only if $E\left[\left(\frac{1}{p_s} - \varepsilon - K_i\right)^+\right]^2 < \infty$, where $\left(\frac{1}{p_s} - \varepsilon - K_i\right)^+ = \max(\frac{1}{p_s} - \varepsilon - K_i, 0)$.

The first and second moments of $\left(\frac{1}{p_s} - \varepsilon - K_i\right)$ are:

$$E\left(\frac{1}{p_s} - \varepsilon - K_i\right) = \frac{1}{p_s} - \varepsilon - \frac{1}{p_s} = -\varepsilon \qquad (12)$$

$$E\left[\left(\left(\frac{1}{p_s} - \varepsilon - K_i\right)^+\right)^2\right] \leq E\left(\frac{1}{p_s} - \varepsilon - K_i\right)^2 = \left(\frac{1}{p_s} - \varepsilon\right)^2 - 2\left(\frac{1}{p_s} - \varepsilon\right)\frac{1}{p_s} + E(K_i^2)$$

$$= \frac{2 - p_s}{p_s^2} + \left(\frac{1}{p_s} - \varepsilon\right)^2 - 2\left(\frac{1}{p_s} - \varepsilon\right) < \infty. \qquad (13)$$

As a result, we have

$$E\left[\limsup_{n\to\infty}\left(\sum_{i=1}^n \left(\frac{1}{p_s} - \varepsilon - K_i\right)\right)\right] \leq E\left[\sup_{n\geq 0}\sum_{i=1}^n \left(\frac{1}{p_s} - \varepsilon - K_i\right)\right] < \infty \qquad (14)$$

which leads to

$$\limsup_{n\to\infty}\left[\sum_{i=1}^n \left(\frac{1}{p_s} - \varepsilon - K_i\right)\right] < \infty \ a.s. \qquad (15)$$

From (10), (11), and (15), we have $\limsup\limits_{n\to\infty}\left[\frac{T}{2}R_n - \lambda(T + \tau \sum\limits_{i=1}^n K_i)\right] = -\infty \ a.s.$

## APPENDIX III
## PROOF OF THEOREM 1

With samplings at each time observation i.i.d. with each other, a general optimal stopping rule given in [19] is of the form $N^* = \min\{n \geq 1 : Y_n \geq W_n^*\}$ where $W_n^* = E\left(V_{n+1}^* | X(1), ..., X(n)\right)$, $Y_n$ is a reward when stop at $N = n$ and $V_{n+1}^*$ represents expected reward following optimal stopping rule which does not stop before time observation $n+1$. Let $V^*$ represent the maximal average reward. With substitution the general form of $N^*$ can be simplified as $N^* = \min\{n \geq 1 : Y_n \geq V^* - C_n\}$ where $Y_n = \frac{T}{2}R_n - \lambda T - \lambda\tau\sum\limits_{i=1}^n K_i$ indicates the reward and $C_n = \lambda\tau\sum\limits_{i=1}^n K_i$ represents the cost of waiting time until $n$. Further simplifying it, $N^*$ can be described as $N^* = \min\{n \geq 1 : \frac{T}{2}R_n - \lambda T \geq V^*(\lambda)\}$ where $N^*$ is determined by $\lambda$. To derive this rule, $V^*$ needs to be known. By the optimality equation of a general form $V_n^* = \max\{Y_n, E\left(V_{n+1}^* | X(1), ..., X(n)\right)\}$, we have:

$$V^* - \lambda\tau\sum_{i=1}^{n-1} K_i = \max\left\{\frac{T}{2}R_n - \lambda T - \lambda\tau\sum_{i=1}^n K_i, V^* - \lambda\tau\sum_{i=1}^n K_i\right\} \qquad (16)$$

$$V^* = \max\left\{\frac{T}{2}R_n - \lambda T, V^*\right\} - \lambda\tau K_n. \qquad (17)$$

With expectations over two sides, we have:

$$V^*(\lambda) = E\left\{\max\left\{\frac{T}{2}R_n - \lambda T, V^*(\lambda)\right\} - \lambda\tau K_n\right\}. \tag{18}$$

By using [19, Theorem 6.1] if $\sup\limits_{N\geq 0}\left\{\frac{T}{2}E(R_N) - \lambda E\left(T + \tau\sum\limits_{i=1}^{N} K_i\right)\right\} = V^*(\lambda) = 0$ is achieved by a rule $N^*$, it also achieves $\sup\limits_{N\geq 0}\frac{E(Y_N)}{E(T_N)}$. By setting $V^*(\lambda^*) = 0$, we can derive $\lambda^*$ as the solution of $E\left\{\max\left\{\frac{T}{2}R_n - \lambda T, 0\right\}\right\} = \lambda\tau E(K_n) = \frac{\lambda\tau}{p_s}$, which is the maximal system throughput. As a result, the optimal stopping rule turns out to be $N^* = \min\{n \geq 1 : R_n \geq 2\lambda^*\}$.

## APPENDIX IV
## PROOF OF THEOREM 2

We have

$$E\left\{\max\left\{\frac{T}{2}R_n - \lambda T, 0\right\}\right\} = \int_{\lambda T}^{+\infty}(x - \lambda T)\mathrm{d}F_{\frac{T}{2}R_n}(x)$$

where $F_{\frac{T}{2}R_n}(x)$ is the cumulative distribution function (CDF) of $\frac{T}{2}R_n$. Therefore, we can say that $E\left\{\max\left\{\frac{T}{2}R_n - \lambda^*T, 0\right\}\right\}$ is continuous and a decreasing function with respect to $\lambda$. Also, $\frac{\lambda\tau}{p_s}$ is a strictly increasing function with respect to $\lambda$. Both functions are positive. Then it can seen that they two functions have one and only one intersection point in a two-dimensional plot with the horizontal axis being $\lambda$.

## APPENDIX V
## PROOF OF THEOREM 3

To guarantee the existence of the optimal stopping rule in the sub-layer problem, two sufficient conditions need to be proved. As a basis, we first prove the finiteness of $E(R_m)$ where $m$ is the index of observation in the sub-layer stopping problem.

$$\begin{aligned}
E(R_m) &= E\left\{\sum_{j=1}^{L}\left\{I([s(m)=j])\left\{\log_2(1 + \frac{P_sP_r|f_{s(n)j}(n)|^2|g_{js(n)}(m)|^2}{1+P_s|f_{s(n)j}(n)|^2+P_r|g_{js(n)}(m)|^2})\right\}\right\}\right\} \\
&= \sum_{j=1}^{L}\left\{\frac{1}{L}E\left\{\log_2(1 + \frac{P_sP_r|f_{s(n)j}(n)|^2|g_{js(n)}(m)|^2}{1+P_s|f_{s(n)j}(n)|^2+P_r|g_{js(n)}(m)|^2})\right\}\right\} \\
&\overset{(b)}{\leq} \sum_{j=1}^{L}\left\{\frac{1}{L}\frac{1}{\ln 2}P_sP_r|f_{s(n)j}(n)|^2 E(|g_{js(n)}|^2)\right\} \\
&= \sum_{j=1}^{L}\left\{\frac{1}{L}\frac{1}{\ln 2}P_sP_r|f_{s(n)j}(n)|^2 \sigma_g^2\right\} < \infty
\end{aligned} \tag{19}$$

where $(b)$ comes from (9).

With the finite property of $E(R_m)$ (which is expectation over channel gains of the second hop), similar to the proofs of Lemma 1 and Lemma 2, the sufficient conditions of the optimal stopping rule in the sub-layer can be proved. With the reward modified to $Y_m = \frac{T}{2}R_m - \lambda\frac{T}{2} - \lambda\frac{\tau}{2}\sum_{i=1}^{m}K_i$, by following the way in Appendix III we can obtain an optimal stopping rule for the sub-layer as the form: $M^* = \min\{m \geq 1 : R_m \geq \lambda^*\}$ where $\lambda^*$ satisfies the equality $E\{\max\{R_m - \lambda, 0\}\} = \frac{\lambda\tau}{Tp_r}$. Similar to Appendix IV, the existence and uniqueness of $\lambda^*$ can be easily derived.

## APPENDIX VI
## PROOF OF COROLLARY 3

With $\{f_{s(n)1}(n), ..., f_{s(n)L}(n)\}$ sampled in the main-layer problem, we check properties of the roots $\lambda^*$ of $E\{\max\{R_m - \lambda, 0\}\} = \frac{\lambda\tau}{Tp_r}$. At first, we can find that $E\{\max\{R_m - \lambda, 0\}\}$ is a decreasing function with respect to $\lambda$ and $\frac{\lambda\tau}{Tp_r}$ is increasing function with respect to $\lambda$. Hence, the uniqueness and non-negativeness of the root $\lambda^*$ is guaranteed.

Further, we have
$$\begin{array}{rl}\frac{\lambda^*\tau}{Tp_r} = & E\{\max\{R_m - \lambda^*, 0\}\} \\ \leq & E\{R_m\} \\ \text{from (19)} < & \infty \end{array} \quad (20)$$

which leads to $\lambda^* < \infty$.

Stopping time $M$ in the sub-layer is geometrically distributed and then according to wald theorem [19], $E(T_{M^*})$ is finite. With $\lambda^*$ finite, we can also derive the finiteness of $E(Y_{M^*})$.

## APPENDIX VII
## PROOF OF THEOREM 4

In a similar way as the proof of Appendix III, we need to find an optimal stopping rule to maximize $E\left\{\lambda^*E(T_{M^*}) - \gamma\left(E(T_{M^*}) + \frac{T}{2} + \frac{\tau}{2}\sum_{i=1}^{N}K_i\right)\right\}$, where $\gamma$ can be viewed as the system throughput. As functions of samplings

$$\{f_{s(n)1}(n), ..., f_{s(n)L}(n)\}$$

at observation $n$ in the main-layer problem, $\lambda^*E(T_{M^*})$ and $E(T_{M^*})$ can be respectively denoted as $R_n^1$ and $R_n^2$. As a result, the problem changes to how to find an stopping rule $N^*$ which

satisfies

$$E\left\{R^1_{N^*} - \gamma\left(R^2_{N^*} + \frac{T}{2} + \frac{\tau}{2}\sum_{i=1}^{N^*} K_i\right)\right\} = \sup_{N \geq 0}\left\{E\left\{R^1_N - \gamma\left(R^2_N + \frac{T}{2} + \frac{\tau}{2}\sum_{i=1}^{N} K_i\right)\right\}\right\} = 0$$

where $R^1_n \geq 0$, $R^2_n \geq 0$ and $\gamma \geq 0$. Based on Lemma 2, it is clear that

$$\limsup_{n \to \infty}\left\{R^1_n - \gamma\left(R^2_n + \frac{T}{2} + \frac{\tau}{2}\sum_{i=1}^{n} K_i\right)\right\} = -\infty.$$

The other condition for optimal stopping rule's existence can be proved as follows:

$$E\left\{\sup_n\left\{R^1_n - \gamma\left(R^2_n + \frac{T}{2} + \frac{\tau}{2}\sum_{i=1}^{n} K_i\right)\right\}\right\} \leq E\left\{\sup_n\left\{R^1_n - n\gamma\frac{\tau}{2}\left(\frac{1}{p_s} - \varepsilon\right)\right\}\right\} +$$

$$E\left\{\sup_n\left\{\gamma\frac{\tau}{2}\sum_{i=1}^{n}\left(\frac{1}{p_s} - \varepsilon - K_i\right)\right\}\right\} - \gamma\frac{T}{2} + E\left\{\sup_n\left\{-\gamma R^2_n\right\}\right\} \quad (21)$$

where $0 < \varepsilon < \frac{1}{p_s}$. With $R^1_n$ finitely valued, we have $E(R^1_n) < \infty$. By using Theorem 4.1 and 4.2 in [19], we can prove finiteness of the first and second terms on the right-hand side of inequality (21). Together with the fact that $E\left\{\sup_n\{-\gamma R^2_n\}\right\} < 0$, we have

$$E\left\{\sup_n\left\{R^1_n - \gamma\left(R^2_n + \frac{T}{2} + \frac{\tau}{2}\sum_{i=1}^{n} K_i\right)\right\}\right\} < \infty.$$

After proofs of sufficient conditions, we can find an optimal stopping rule $N^*$. By the optimality equation, we obtain:

$$V^* - \gamma\frac{\tau}{2}\sum_{i=1}^{n-1} K_i = \max\left\{R^1_n - \gamma\frac{T}{2} - \gamma R^2_n - \gamma\frac{\tau}{2}\sum_{i=1}^{n} K_i, V^* - \gamma\frac{\tau}{2}\sum_{i=1}^{n} K_i\right\} \quad (22)$$

$$V^* = \max\left\{R^1_n - \gamma\frac{T}{2} - \gamma R^2_n, V^*\right\} - \gamma\frac{\tau}{2}K_n. \quad (23)$$

Similar to the way in Appendix III, we finally acquire the optimal stopping rule which maximizes the relay system throughput as the form $N^* = \min\left\{n \geq 1 : R^1_n - \gamma^* R^2_n \geq \gamma^*\frac{T}{2}\right\}$ where $\gamma^*$ solves the equation $E\left\{\max\left\{R^1_n - \gamma R^2_n - \gamma\frac{T}{2}, 0\right\}\right\} = \frac{\gamma\tau}{2p_s}$.

## APPENDIX VIII
## PROOF OF THEOREM 5

For fixed $\gamma > 0$, we need to find an optimal stopping rule to maximize $E(Y_M - \gamma T_M)$ which is conditioned on a channel gain realization in the first hop. The sufficient conditions for existence of an optimal stopping rule can be proved similar to Appendix VI. Different from the sub-layer in the intuitive stopping rule (which maximizes the sub-layer system throughput), here we maximize $E(Y_M - \gamma T_M)$, and thus have the optimal stopping rule: $M(\gamma) = \min\left\{m \geq 1 : \frac{T}{2}R_m \geq W^*(\gamma) + \frac{T}{2}\gamma\right\}$ where $W^*(\gamma)$ satisfies the optimality equation

$$E\left[\max\left(\frac{T}{2}R_m - \frac{T}{2}\gamma, W^*(\gamma)\right)\right] = W^*(\gamma) + \frac{\gamma\tau}{2p_r}.$$

Since $E\left[\max\left(\frac{T}{2}R_m - \frac{T}{2}\gamma, W^*(\gamma)\right)\right]$ is a continuously decreasing function and $W^*(\gamma) + \frac{\gamma\tau}{2p_r}$ is an increasing function, a finitely unique intersection point $W^*(\gamma)$ should always exist.

## APPENDIX IX
## PROOF OF THEOREM 6

Before deriving the main-layer optimal stopping rule $N^*$, an extended version of Theorem 6.1 in [19] should be given as a proof basic. That is, if $\sup_{N \geq 0} E[Y_N(\gamma) - \gamma T_N] = 0$ for fixed $\gamma$ we should have $E[Y_N(\gamma) - \gamma T_N] \leq 0$ which leads to $\frac{E(Y_N(\gamma))}{E(T_N)} \leq \gamma$.

With $N^*$ such that $E[Y_{N^*}(\gamma) - \gamma T_{N^*}] = 0$, it can also achieve $\sup_{N \geq 0} \frac{E[Y_N(\gamma)]}{E[T_N]}$. With the finite property of $W^*(\gamma)$ which is a function of $\{s(n), f_{s(n)1}(n), ..., f_{s(n)L}(n)\}$ proved before, we can get $E[W^*(\gamma)] < \infty$. In a similar way to Appendix I and II, the conditions which ensure the existence of the optimal stopping rule hold. Based on that, we can get the optimal stopping rule $N(\gamma) = \{n \geq 1 : W^*(\gamma) - \frac{T}{2}\gamma \geq V^*(\gamma)\}$ where $V^*(\gamma)$ satisfies the optimality equation as

$$E\left[\max(W^*(\gamma) - \frac{T}{2}\gamma, V^*(\gamma))\right] = V^*(\gamma) + \frac{\tau\gamma}{2p_s}.$$

By setting $V^*(\gamma) = 0$, the maximal system throughput $\gamma^*$ is the solution of

$$E\left[\max(W^*(\gamma) - \frac{T}{2}\gamma, 0)\right] = \frac{\tau\gamma}{2p_s}.$$

Therefore, the optimal stopping rule achieving $\gamma^*$ is of the form: $N^* = \left\{n \geq 1 : W^*(\gamma^*) \geq \frac{T}{2}\gamma^*\right\}$.